\pdfoutput=1
\documentclass[copyright,creativecommons]{eptcs}
\providecommand{\event}{ICLP 2026}

\usepackage{amsmath}
\usepackage{amssymb,amsfonts}
\usepackage{mathtools}
\usepackage{bm}
\usepackage{relsize}

\usepackage{xcolor}

\usepackage{graphicx}

\usepackage{xspace}
\usepackage{cleveref}

\usepackage{algorithm}
\usepackage[noend]{algpseudocode}

\usepackage{multicol}
\usepackage{multirow}
\usepackage{verbatimbox}
\usepackage{wrapfig}
\usepackage{fancyvrb}

\usepackage{enumitem}

\usepackage{amsthm}
\theoremstyle{definition}
\newtheorem{definition}{Definition}[section]
\newtheorem{example}[definition]{Example}

\theoremstyle{plain}

\theoremstyle{remark}
\newtheorem{remark}[definition]{Remark}

\usepackage{thmtools,thm-restate}

\newif\ifappendix
\appendixfalse
\ifappendix
  \newcommand{\appref}[2]{Appendix~\ref{#1}}
\else
  \newcommand{\appref}[2]{#2}
\fi

\newcommand{\arxivref}{the full paper~\cite{shapiro2025glp}}

\newenvironment{keywords}{\par\smallskip\noindent\textbf{Keywords:} }{}

\setlist{nosep, leftmargin=*}
\setlist{itemsep=1pt, topsep=3pt, leftmargin=*}

\usepackage{etoolbox}
\usepackage{titlesec}
\makeatletter
\def\thm@space@setup{\thm@preskip=4pt \thm@postskip=4pt}
\makeatother
\BeforeBeginEnvironment{verbatim}{\vspace{-0.3\baselineskip}}
\AfterEndEnvironment{verbatim}{\vspace{-0.4\baselineskip}}
\titlespacing*{\section}{0pt}{8pt}{4pt}
\titlespacing*{\subsection}{0pt}{6pt}{3pt}
\titlespacing*{\subsubsection}{0pt}{4pt}{2pt}

\newcommand{\mypara}[1]{\vspace{1pt}\noindent\textbf{#1.}}
\newcommand{\temph}[1]{\emph{#1}}

\newcommand{\remove}[1]{}

\newcommand{\calV}{\mathcal{V}}
\newcommand{\calG}{\mathcal{G}}
\newcommand{\calT}{\mathcal{T}}

\newcommand{\calC}{\mathcal{C}}

\newcommand{\ia}{\textit{i}}
\newcommand{\ib}{\textit{ii}}

\newcounter{pc}

\providecommand{\titlerunning}{Grassroots Logic Programs}
\providecommand{\authorrunning}{Shapiro}

\providecommand{\thisvolume}[1]{this volume of EPTCS, Open Publishing Association}

\title{GLP: A Grassroots, Multiagent, Concurrent, Logic Programming Language for AI}

\author{Ehud Shapiro
\institute{London School of Economics and Weizmann Institute of Science}
}

\begin{document}

\maketitle

\begin{abstract}
A grassroots platform is a multiagent distributed system in which multiple independent instances can form and operate independently of each other and of any global resource, yet may coalesce into ever larger instances, possibly resulting in a single global instance. Grassroots platforms aim to offer an egalitarian/democratic alternative to centralised/autocratic and decentralised/plutocratic global platforms.

Here, we present Grassroots Logic Programs (GLP), a multiagent concurrent logic programming language designed for the implementation of grassroots platforms: we recall the standard operational semantics of logic programs; introduce the concurrent operational semantics of GLP as its restriction; recall multiagent atomic transactions; use them to introduce a multiagent operational semantics of GLP; and prove multiagent GLP to be grassroots.

These mathematical foundations are being used by AI to implement GLP as well as to program in GLP: a workstation-based implementation of concurrent GLP in Dart was derived from the concurrent operational semantics of GLP; a multiagent smartphone-based implementation of GLP in Dart/Flutter is being developed based on the multiagent operational semantics of GLP; a moded type system for GLP was designed to facilitate collaborative human-AI development of GLP programs; GLP implementations of grassroots platforms for the social graph, social networks, currencies and bonds have been derived by AI from mathematical specifications.

While concurrent logic programming and its powerful programming techniques have been known for four decades, adoption has been hampered by their inaccessibility to the average programmer. With AI as a hyper-programmer, this limitation is removed and the abstract nature and expressive power of concurrent logic programming can be put into effective use.
\end{abstract}


% Main sections
\section{Introduction}

\mypara{Grassroots} Grassroots platforms aim to offer an egalitarian and democratic alternative to centralised and autocratic (Facebook) and decentralised and plutocratic (Bitcoin) global platforms~\cite{shapiro2025characterising}. A digital platform is \emph{grassroots}~\cite{shapiro2023grassrootsBA,shapiro2025atomic} if it can have multiple instances that can (\ia)~operate independently of each other and of any global resource other than the network, and (\ib)~coalesce into ever larger instances, possibly resulting in a single global instance.  A grassroots platform aims to operate solely on the smartphones of its participants.

Specifications of grassroots platforms that have been provided and proven grassroots include the grassroots social graph~\cite{shapiro2023gsn,shapiro2025atomic} --- the infrastructure platform on which all others build --- grassroots social networks~\cite{shapiro2023gsn,shapiro2026cssn}, grassroots cryptocurrencies and bonds~\cite{shapiro2024gc,lewis2023grassroots,shapiro2026bonds}, and grassroots federations~\cite{shapiro2025GF,halpern2024federated}; the unifying formal framework being volitional multiagent atomic transactions~\cite{lewis2026volitional}. Working GLP implementations of the social graph (this paper), child-safe social networks~\cite{shapiro2026cssn}, grassroots cryptocurrencies and bonds~\cite{shapiro2026bonds}, and secure recovery from major faults~\cite{eitan2026securing} have been derived by AI from these specifications. The Scuttlebutt protocol and social network~\cite{kermarrec2020gossiping} is perhaps the sole example of a deployed grassroots platform to date.

\mypara{Grassroots Logic Programs} Here, we present Grassroots Logic Programs (GLP), a multiagent, concurrent, logic programming language, designed for the implementation of grassroots platforms. Syntactically, GLP extends Logic Programs (LP)~\cite{lloyd1987foundations} with:
\begin{enumerate}
    \item \textbf{Readers:} Each logic variable $X$ (now referred to as \emph{writer}) is paired with a \emph{reader} $X?$, which is assigned a value only once its paired writer $X$ is assigned that value; 
    \item  \textbf{Single-Occurrence (SO):} A variable may occur at most once in a goal or a clause; and
    \item \textbf{Single-Reader Single-Writer (SRSW):} A writer occurs in a clause iff its paired reader does.
\end{enumerate}
The result eschews unification for simple term matching, and conjures both linear logic~\cite{girard1987linear} and futures/promises~\cite{baker1977future,friedman1976impact}: An assignment to a variable may be produced at most once, via the sole occurrence of a writer (promise), and consumed at most once, via the sole occurrence of its paired reader (future).  

Hence, in a multiagent distributed implementation, if a writer and its paired reader are held by different agents, an assignment to the writer is realised as a single message from the writer-holding agent to the reader-holding agent.  The source of all the powerful concurrent logic programming techniques is that such a message may, in turn, contain further readers and writers, enabling the concise expression of rich multidirectional communication modalities.  In the simplest case, a reader sent in a message keeps the channel open for further messages from the sender (streaming); a writer sent in a message allows the receiver to reply.  More generally, variables in a message paired with variables held by third agents can be used for arbitrary network reconfiguration (e.g. friend-mediated introduction, example below).

\mypara{Semantics} The operational semantics of GLP is presented in two stages: First, we present the \emph{concurrent operational semantics of GLP} (cGLP), an interleaving-based concurrent (single-agent) semantics defined as a restriction of standard LP semantics, preserving computation-as-deduction~\cite{kowalski1974predicate}.

Second, we present multiagent transition systems and their specification via multiagent atomic transactions, and use them to define the \emph{multiagent operational semantics of GLP} (maGLP), with (1) \textbf{Communication:}  Named agents that operate independently while communicating to remote readers assignments made to paired local writers;  (2) \textbf{Cold-calls:} A means for sending a term with variables to a named agent while retaining their paired variables locally.
Cold-calls allow two agents in disconnected components of the social graph to become the owners of paired logic variables. We illustrate maGLP via the grassroots social graph, in which agents may initiate friendships (bidirectional communication channels) via ``cold-calls'' as well as introduce mutual friends to each other.
Lastly, we recall the definition of grassroots platforms and how to prove that a platform specified via atomic transactions is grassroots~\cite{shapiro2023grassrootsBA,shapiro2025atomic}, and apply these to prove that maGLP is indeed grassroots.

\mypara{Historical Context} The concurrent logic programming family emerged in the 1980s --- Concurrent Prolog~\cite{shapiro1983subset}, GHC~\cite{ueda1986guarded}, PARLOG~\cite{clark1986parlog}; surveyed in~\cite{shapiro1989family} --- and was simplified by \emph{flattening} (restricting guards to primitive tests) in Flat Concurrent Prolog (FCP)~\cite{mierowsky1985fcp} and Flat GHC; sequential FCP abstract machines~\cite{houri1989sequential} made commercial deployment feasible. Mode systems followed, including Ueda's moded Flat GHC~\cite{ueda1994moded,ueda1995io,ueda2001resource}, anticipating GLP's SO discipline as a type-system refinement. GLP can be understood as FCP with the SRSW restriction added, simplifying read-only unification~\cite{levi1985readonly}. These languages saw industrial deployment during the Japanese Fifth Generation project~\cite{moto1983overview,shapiro1983fifth}, most prominently in the chat product Virtual Places, developed in FCP by Ubique~\cite{Ubique} and acquired by AOL in 1995, after which the codebase was migrated to C. With the conclusion of the Fifth Generation project~\cite{shapiro19935th}, concurrent logic programming went out of fashion. Its core ideas, however, persisted: pattern-matched message-passing among lightweight share-nothing processes, with single-assignment variables for synchronisation, found their most prominent industrial expression in Erlang~\cite{armstrong2010erlang} --- perhaps the closest descendant of concurrent logic programming --- and from there to Elixir and to actor frameworks such as Akka~\cite{akka2022} in Scala and Orleans~\cite{orleans2022} in .Net. What was lost in this evolution was logic programming's metaprogramming~\cite{safra1988meta,lichtenstein1988concurrent} and the seamless integration of computation and synchronisation through paired logical variables; GLP recovers these, while maintaining the simplicity and efficiency of futures/promises.

\mypara{AI} The foundations presented here are used by AI (Claude) in two distinct disciplines. \emph{Implementing} GLP proceeds through three layers---a mathematical specification, an informal English-and-code specification derived from it by AI, and Dart code derived from that---with authority flowing math$\to$spec$\to$Dart but harmonised by back-and-forth. The concurrent and multiagent operational semantics are first refined into deterministic counterparts, dGLP and madGLP~\cite{shapiro2026implementing}. From dGLP AI derived a workstation-based GLP implementation in Dart, and from madGLP a smartphone-based multiagent one in Dart/Flutter is being developed. Running these implementations surfaced defects, several at the mathematical level, that drove revisions to the operational semantics themselves. A moded type checker was implemented by AI in the same way~\cite{shapiro2026types}. \emph{Programming} in GLP proceeds through three layers of its own---mathematical specifications, provided as guarded multiagent atomic transactions~\cite{lewis2026volitional}; the type definitions and declarations agreed for the program together with the intended behaviour of each procedure; and the GLP code: the designer and AI first agree the types and intent, and only then does AI write, type-check, test, and debug the code, with the human providing design-level oversight rather than code-level intervention~\cite{shapiro2026types}. The type checker catches at compile time a characteristic class of mode errors---confusing a reader for its paired writer---otherwise manifesting as silent run-time suspensions; a befriending clause, for example, was rejected because a reader argument received a produced rather than a consumed value, and AI corrected it from the diagnostic alone. A secure GLP implementation using public-key cryptography, mutual attestations, and friend-based identity custodians is also being developed~\cite{eitan2026securing}. Example GLP programs in this paper are typed.

\mypara{Paper outline}
Section~\ref{sec:glp} presents GLP. Section~\ref{sec:maglp} presents multiagent transition systems, the maGLP definition, and its safety properties. Section~\ref{sec:ma-social-graph} presents the grassroots social graph. Section~\ref{sec:grassroots} proves maGLP is grassroots. Section~\ref{section:conclusion} concludes. \ifappendix All proofs are deferred to the appendices.\else Proofs and supporting material are in  \arxivref.\fi

%==============================================================================
\section{GLP}
\label{sec:glp}
%==============================================================================

We present GLP syntax and operational semantics and discuss its programming techniques.

%------------------------------------------------------------------------------
\subsection{Syntax}
\label{sec:glp-ext}
%------------------------------------------------------------------------------

GLP extends LP by adding a paired \emph{reader} $X?$ to every ``ordinary'' logic variable $X$, now called a \emph{writer}.

\begin{definition}[GLP Variables]
\label{def:glp-variables}
Let $\calV$ denote the set of LP variables (identifiers beginning with uppercase), henceforth called \temph{writers}. Define $\calV? = \{X? \mid X \in \calV\}$, called \temph{readers}. The set of all GLP variables is $\hat\calV = \calV \cup \calV?$. A writer $X$ and its reader $X?$ form a \temph{variable pair}.
\end{definition}
GLP terms, unit goals, goals, and clauses are as in LP but defined over the variables in $\hat\calV$.

\begin{definition}[Single-Occurrence (SO) Invariant]
\label{def:so-invariant}
A term, goal, or clause satisfies the \temph{single-occurrence (SO) invariant} if every variable occurs in it at most once.
\end{definition}

\begin{definition}[GLP Program, Goals]
\label{def:glp-program}
A clause $C$ satisfies the \temph{single-reader/single-writer (SRSW) restriction} if it satisfies SO and a variable occurs in $C$ iff its paired variable also occurs in $C$.
A \temph{GLP program} is a finite sequence of clauses satisfying SRSW; clauses for the same predicate form a \temph{procedure}.
The set of GLP goals $\hat\calG(P)$ includes all goals over $\hat\calV$ and the vocabulary of $P$ that satisfy SO.
\end{definition}
The purpose of the SRSW restriction is to prevent multiple writer occurrences racing to assign a variable; the simpler (SW) restriction is insufficient, since if a reader with multiple occurrences is assigned a term with a writer, the result is a writer with multiple occurrences.
We use set notation also when referring to multisets.
\begin{example}[Fair Merge]
\label{ex:merge}
Consider the quintessential concurrent logic program for fairly merging two streams, written in GLP:\footnote{Moded-type definitions (\texttt{T ::= ...}) and declarations (\texttt{procedure ...}) used informally in examples are formally introduced in the Typed GLP companion paper~\cite{shapiro2026types}.}
\begin{verbatim}
Stream ::= [] ;[_|Stream].

procedure merge(Stream?, Stream?, Stream).
merge([X|Xs], Ys, [X?|Zs?]) :- merge(Ys?, Xs?, Zs).
merge(Xs, [Y|Ys], [Y?|Zs?]) :- merge(Xs?, Ys?, Zs).
merge([], Ys, Ys?).
merge(Xs, [], Xs?).
\end{verbatim}
and the goal \verb=merge([1,2,3|Xs?],[a,b|Ys?],Zs)=, which satisfies SO; each clause satisfies SRSW. The first clause swaps inputs in the recursive call, ensuring fairness when both streams are available.
\end{example}

%------------------------------------------------------------------------------
\subsection{Operational Semantics}
\label{sec:glp-operational}
%------------------------------------------------------------------------------

\begin{definition}[Transition System, Computation, Run, Safe, Live, Correct (cf.~\cite{shapiro2021multiagent,lewis2026volitional})]
\label{def:ts}
A \temph{transition system} is a tuple $TS = (C, c_0, T, {\sim})$ where $C$ is an arbitrary set of \temph{configurations}, $c_0 \in C$ a designated \temph{initial configuration}, $T \subseteq C \times C$ a set of \temph{transitions}, each a pair $c \rightarrow c'$ of non-identical configurations $c \ne c' \in C$, and ${\sim}$ an equivalence relation on $T$; we write $[t]$ for the class of $t \in T$ under ${\sim}$.

A \temph{computation} is a (nonempty, finite or infinite) sequence of configurations $c_1, c_2, \ldots$; it is a \temph{run} if $c_1 = c_0$, and \temph{safe} if $c_i \rightarrow c_{i+1} \in T$ for every two consecutive configurations. We write $c \xrightarrow{*} c'$ for the existence of a safe computation from $c$ to $c'$ (empty if $c = c'$). A class $[t] \in T/{\sim}$ is \temph{enabled} in $c$ if $c \rightarrow c' \in [t]$ for some $c'$. A run is \temph{live} if no class $[t]$ is enabled in every configuration of some suffix in which no member of $[t]$ occurs, and \temph{correct} if it is safe and live.
\end{definition}

\begin{definition}[Substitutions and Assignments]
\label{def:writers-assignment}
A GLP \temph{writer assignment} is a term of the form $X := T$, $X\in\calV$, $T\notin\calV$, satisfying SO. Similarly, a GLP \temph{reader assignment} is a term of the form $X? := T$, $X?\in\calV?$, $T\notin\calV$, satisfying SO. A \temph{writers (readers) substitution} $\sigma$ is the substitution implied by a set of writer (reader) assignments that jointly satisfy SO. Given a writers assignment $X := T$, its \temph{readers counterpart} is $X? := T$, and given a writers substitution $\sigma$, its \temph{readers counterpart} $\sigma?$ is the readers substitution defined by $X?\sigma? = X\sigma$. Given a reader assignment $X? := T$, its \temph{writers counterpart} is $X := T$, and given a readers substitution $\tau$, its \temph{writers counterpart} $\tau!$ is the writers substitution defined by $X\tau! = X?\tau$. The \temph{pair completion} of a readers substitution $\sigma$ is $\sigma^\star = \sigma \cup \sigma!$, applied to a fixed point.
\end{definition}

\begin{definition}[GLP Renaming, Renaming Apart]
\label{def:glp-renaming}
A \temph{GLP renaming} is an injective substitution $\rho: \hat\calV \to \hat\calV$ such that for each $X \in \calV$: $X\rho \in \calV$ and $X?\rho = (X\rho)?$. Two GLP terms \temph{have a variable in common} if for some writer $X \in \calV$, either $X$ or $X?$ occurs in both. A GLP renaming $\sigma$ \temph{renames $T'$ apart from} $T$ if $T'\sigma$ and $T$ have no variable in common.
\end{definition}

\begin{definition}[Writer MGU]
\label{def:writer-mgu}
Given two GLP unit goals $A$ and $H$, a \temph{writer mgu} is a writers substitution $\sigma$ such that $A\sigma = H\sigma$ and $\sigma$ is most general among such substitutions. 
\end{definition}

\begin{definition}[GLP Goal/Clause Reduction]
\label{def:glp-reduction}
Given GLP unit goal $A$ and clause $C$, with $H$ \verb|:-| $B$ being the result of the GLP renaming of $C$ apart from $A$, the \temph{GLP reduction} of $A$ with $C$ \temph{succeeds with result} $(B,\sigma)$ if $A$ and $H$ have a writer mgu.
\end{definition}

The cGLP transition equivalence below identifies ``the same transition'' across configurations as goals are instantiated by reader substitutions; following~\cite{lewis2026volitional}, we first assign each goal a persistent identity.

\begin{definition}[Goal Identity]
\label{def:goal-identity}
Each unit goal in a cGLP computation carries a unique \temph{identifier} assigned at spawn time: goals in the initial goal $G_0$ receive distinct initial identifiers, and goals spawned by a Reduce transition receive fresh identifiers. Communicate transitions preserve identifiers: the goal containing $X?$ retains its identity after instantiation.
\end{definition}

\begin{definition}[cGLP Transition System]
\label{def:cglp-ts}
Given a GLP program $P$, an \temph{asynchronous resolvent} over $P$ is a pair $(G, \sigma)$ where $G \in \hat\calG(P)$ and $\sigma$ is a readers substitution.

A transition system $cGLP = (\calC, c_0, \calT, {\sim})$ is a \temph{cGLP transition system} over $P$ and initial goal $G_0$ satisfying SO if:
\begin{enumerate}
    \item $\calC$ is the set of all asynchronous resolvents over $P$
    \item $c_0 = (G_0, \emptyset)$
    \item $\calT$ is the set of all transitions $(G, \sigma) \rightarrow (G', \sigma')$ satisfying either:
    \begin{enumerate}
        \item \textbf{Reduce:} there exists unit goal $A \in G$ such that $C \in P$ is the first clause for which the GLP reduction of $A$ with $C$ succeeds with result $(B, \hat\sigma)$, $G' = (G \setminus \{A\} \cup B)\hat\sigma$, and $\sigma' = \sigma \circ \hat\sigma?$
        \item \textbf{Communicate:} $\{X? := T\} \in \sigma$, $X?\in G$, $G' = G\{X? := T\}$, and $\sigma' = \sigma$
    \end{enumerate}
    \item ${\sim}$, the \temph{cGLP transition equivalence}, relates $t_1 \sim t_2$ iff either both are Reduce transitions reducing the same goal (by identity, Definition~\ref{def:goal-identity}) with the same clause, or both are Communicate transitions applying the counterpart of the same writer assignment to the same goal (by identity)
\end{enumerate}
\end{definition}

cGLP Reduce differs from LP in (1) the use of a writer mgu instead of a regular mgu and (2) the choice of the first applicable clause instead of any clause. The first is the fundamental use of GLP readers for communication and synchronisation. The second compromises on the or-nondeterminism of LP to allow writing fair concurrent programs, such as fair merge above. Note that or-nondeterminism is not completely eliminated, as different scheduling of arrival of assignments on the two input streams of \verb|merge| may result in different orders in its output stream.

The cGLP Communicate rule realises the use of reader/writer pairs for asynchronous communication: it communicates an assignment from its writer to its paired reader.

\mypara{Monotonicity}
In LP, if a goal cannot be reduced, it will never be reduced. In cGLP, a goal that cannot be reduced now may be reduced in the future: if $A$ and $H$ have an mgu that writes on a reader $X? \in A$ with no writer mgu at present,  another goal with $X$ may reduce, assigning $X$, and later $X?$, to a value that will allow $A$ and $H$ to have a writer mgu. Conversely, in LP, if a goal $A$ can be reduced now with some clause $H$\verb|:-|$B$, with a regular mgu of $A$ and $H$, it may not be reducible in the future due to variables that $A$ shares with other goals being assigned values by reductions of other goals, preventing unification between the instantiated $A$ and $H$. In cGLP, if a goal $A$ can be reduced now (with a writers mgu), it can always be reduced in the future, as the SO invariant ensures that no other goal can assign any writer in $A$.

Implementation-wise, if a GLP goal $A$ cannot be reduced now, but there is a readers substitution $\sigma$ such that $A\sigma$ can be reduced, such readers are identified, the goal $A$ \emph{suspends} on these readers, and is rescheduled for another reduction attempt once any of them is assigned.

Despite these differences, cGLP has the same notion of logical consequence as LP.

\begin{definition}[cGLP Proper Run, Outcome]
\label{def:cglp-proper-run}
A cGLP run $\rho: (G_0,\sigma_0) \rightarrow \cdots \rightarrow (G_n, \sigma_n)$ is \temph{proper} if for any $1\le i< n$, a variable that occurs in $G_{i+1}$ but not in $G_i$ also does not occur in any $G_j$, $j<i$; the \temph{outcome} of a proper run is $(G_0$ \texttt{:-} $G_n)\sigma_n^\star$; and the run is \temph{successful} if $G_n=\emptyset$.
\end{definition}
Pair completion (Definition~\ref{def:writers-assignment}) is employed since $\sigma_n$ accumulates only readers counterparts, leaving the writers of $G_0$ unbound. Let $/?$ be an operator that replaces every reader by its paired writer.

\begin{restatable}[cGLP Computation is Deduction]{proposition}{propGLPDeduction}
\label{prop:cglp-deduction}
Let $(G_0$ \texttt{:-} $G_n)\sigma_n^\star$ be the outcome of a proper cGLP run $\rho: (G_0,\sigma_0) \rightarrow \cdots \rightarrow (G_n, \sigma_n)$. Then $(G_0$ \texttt{:-} $G_n)\sigma_n^\star/?$ is a logical consequence of $P/?$.
\end{restatable}

We note two additional safety properties of cGLP runs.

\begin{restatable}[SO Preservation]{proposition}{propSOPreservation}
\label{prop:so-preservation}
If the initial goal $G_0$ satisfies SO, then every goal in a proper cGLP run satisfies SO.
\end{restatable}

\begin{restatable}[Monotonicity]{proposition}{propMonotonicity}
\label{prop:glp-monotonicity}
In any proper cGLP run, if unit goal $A$ can reduce with clause $C$ at step $i$, then either an instance of $A$ has been reduced by step $j > i$, or an instance of $A$ can still reduce with $C$ at step $j$.
\end{restatable}

Every cGLP run is safe, so correctness (Definition~\ref{def:ts}) reduces to liveness.

The SO invariant of GLP allows eschewing unification in favour of \emph{term matching}: if two terms that jointly satisfy SO are unifiable, their mgu maps each variable in one term to a subterm of the other. Term matching thus performs joint term-tree traversal and collects variable assignments along the way; the detailed definition and table appear in \appref{appendix:term-matching}{\arxivref}.

%------------------------------------------------------------------------------
\subsection{Guards}
\label{sec:guards}
%------------------------------------------------------------------------------

GLP clauses may include \emph{guards}---tests that determine clause applicability.

\begin{definition}[Guarded Clause]
\label{def:guarded-clause}
A \temph{guarded clause} has the form $H$ \verb|:-| $G$ \verb"|" $B$, where $H$ is the head, $G$ is a conjunction of guard predicates, and $B$ is the body. The guard separator ``\verb"|"'' is interpreted logically as a conjunction.  Guard arguments are readers paired to head writers.
\end{definition}

Guards have three-valued semantics. Each guard predicate explicitly defines its \emph{success} condition. A guard \emph{suspends} if it does not succeed but some instance of it under a readers substitution would succeed. A guard \emph{fails} if no such instance exists. A guard conjunction succeeds if all members succeed; it suspends if any member suspends and none fail; it fails if any member fails.

Definition~\ref{def:glp-reduction} of a GLP goal/clause reduction is augmented to succeed if the guard also succeeds.

\begin{remark}[Guards and SRSW]
\label{rem:guards-srsw}
Guard occurrences count toward SRSW satisfaction: if $X?$ occurs in a guard, its paired writer $X$ must occur in the head and $X?$ may additionally occur once in the body.
Furthermore, if the success of a guard implies that $X?$ is bound to a ground term, then both $X$ and $X?$ may occur multiple times in the clause. Groundness-implying guards include \verb|ground|, \verb|integer|, \verb|number|, \verb|string|, \verb|constant|, arithmetic comparisons (\verb|<|, \verb|>|, \verb|=<|, \verb|>=|, \verb|=:=|, \verb|=\=|), and ground equality (\verb|=?=|). However,  \verb|known| and \verb|compound| do not imply groundness.
\end{remark}

\begin{remark}[Anonymous Variables]
\label{rem:anonymous-variables}
An \emph{anonymous variable} is any variable whose name begins with \verb|_| (e.g., \verb|_|, \verb|_Out|). Anonymous variables may appear anywhere a writer variable may appear; each occurrence denotes a fresh writer with no paired reader. Anonymous readers (\verb|_?|, \verb|_Name?|) are not permitted.
\end{remark}

\mypara{System predicates and body kernels}
GLP \emph{system predicates} such as \verb|:=| (arithmetic assignment), \verb|now| (clock access), and \verb|=..| (term composition/decomposition) are defined by GLP clauses whose bodies may invoke \emph{body kernel predicates}---runtime-implemented primitives not directly accessible to user programs. The system predicate's own guards ensure that body kernel preconditions are met before invocation. For example, arithmetic assignment is defined recursively as follows:

\begin{verbatim}
Result? := N :- number(N?) | Result = N?.
Result? := X + Y :- number(X?), number(Y?) |
    '_add'(X?, Y?, Result).
Result? := X + Y :- otherwise |
    X1 := X?, Y1 := Y?, Result := X1? + Y1?.
\end{verbatim}
\noindent The base case binds a plain number; the first addition clause guards that both operands are numbers and invokes the \verb|'_add'| body kernel; the fallback clause recursively evaluates subexpressions. The full table of body kernels and the guard reference appear in \appref{appendix:guards-system}{\arxivref}; the complete system-predicate definitions are in \verb|programs/self.glp| of the repository.

\subsection{GLP Programming Techniques}

On the one hand, GLP has no unification, only pattern-matching and single-writer variables. On the other hand, goal reduction may write on several variables atomically. The result is that almost the entire range of concurrent logic programming techniques developed since the 1980s~\cite{shapiro1989family,shapiro1987concurrent} is available in GLP, including: streams and their fair, biased, and self-balancing merge~\cite{shapiro1984fair,shapiro1986multiway}; channels as generalised streams~\cite{tribble1988channels}; messages with reply variables and bounded buffers~\cite{takeuchi1988bounded}; network reconfiguration and distributed programming~\cite{shafrir1988distributed}; object-oriented programming with mutable state via stream-recursion~\cite{shapiro1983object}; systems programming and computation control~\cite{shapiro1984systems,silverman1988logix}; and the full range of meta-interpreters --- fail-safe, termination-detecting, tracing, and algorithmic debugging~\cite{safra1988meta,lichtenstein1988concurrent}. Tested GLP renditions of these and further techniques --- monitors, distributors, network switches, replicators, and the metainterpreter family --- are in the \verb|programs| directory of the public GLP repository.\footnote{\url{https://github.com/EShapiro2/GLP}}

Two techniques require workarounds. The use of logic variables for mutual exclusion --- standard in concurrent logic languages with atomic unification~\cite{shapiro1989family} --- must be achieved in GLP via a monitor~\cite{hoare1974monitors,hansen1973operating}, the standard distributed-computing solution. Similarly, general broadcast --- distributing a value to multiple consumers --- is not directly supported by GLP, since it would require multiple occurrences of the same reader, in violation of SO; it must be achieved by explicit \emph{distributors}.  We note that if moded-type checking can determine that the values to be distributed never contain writers, then the SO restriction on readers may safely be relaxed allowing broadcast natively; a future upgrade of the language.  Excluded is the encoding of logic programs or-parallel by the and-parallelism of Concurrent Prolog~\cite{codish1986compiling,shapiro1989or}, which depends on general unification in a fundamental way.

%==============================================================================
\section{Multiagent GLP}\label{sec:maglp}
%==============================================================================

We extend GLP to multiple agents: Recall the notion of multiagent transition systems via multiagent atomic transactions~\cite{shapiro2021multiagent,shapiro2025atomic};  define multiagent GLP (maGLP) as a transactions-based multiagent transition system; and establish its safety properties. 

%------------------------------------------------------------------------------
\subsection{Multiagent Transition Systems}
\label{sec:mts}
%------------------------------------------------------------------------------
We assume a potentially infinite set of \emph{agents} $\Pi$, but consider only finite subsets of it, so when we refer to a particular set of agents $P \subset \Pi$ we assume $P$ to be nonempty and finite. We use $\subset$ to denote the strict subset relation and $\subseteq$ when equality is also possible.

We use $S^P$ to denote the set $S$ indexed by the set $P$, and if $c\in S^P$ we use $c_p$ to denote the member of $c$ indexed by $p\in P$. Intuitively, think of such a $c\in S^P$ as an array of cells indexed by members of $P$ with cell values in $S$.

\begin{definition}[Local States, Configuration, Transaction, Participants]\label{def:transaction}
Given agents $Q \subset \Pi$ and an arbitrary set $S$ of \temph{local states}, a \temph{configuration} over $Q$ and $S$ is a member of $C:= S^Q$. An \temph{atomic transaction} over \temph{participants} $Q$ and states $S$ is any pair of configurations $t=c\rightarrow c' \in C^2$ such that $c\ne c'$, with $t_p := c_p \rightarrow c'_p$ for any $p\in Q$.
\end{definition}

\begin{definition}[Multiagent Transition System]\label{def:mts}
Given agents $P \subset \Pi$ and an arbitrary set $S$ of \temph{local states} with a designated \temph{initial local state} $s_0\in S$, a \temph{multiagent transition system} over $P$ and $S$ is a transition system $TS= (C,c_0,T,{\sim})$ with \temph{configurations} $C:= S^P$, \temph{initial configuration} $c_0:= \{s_0\}^P$, \temph{transitions} $T\subseteq C^2$ being a set of transactions over $P$ and $S$, and ${\sim}$ an equivalence on $T$.
\end{definition}

Rather than specifying a multiagent transition system over a set of agents $P$ directly, we specify it via atomic transactions.

\begin{definition}[Transaction Closure]\label{def:closure}
Let $P\subset \Pi$, $S$ a set of local states, and $C:=S^P$; for any transition or transaction $t = c\rightarrow c'$, we say $p$ is \temph{stationary} in $t$ if $c_p = c'_p$.
For a transaction $t=(c\rightarrow c')$ over local states $S$ with participants $Q \subseteq P$, the \temph{$P$-closure of $t$}, $t{\uparrow}P$, is the set of transitions over $P$ and $S$ defined by:
$$
t{\uparrow}P := \{ t' \in C^2 :
\forall q\in Q.(t_q = t'_q) \wedge \forall p\in P\setminus Q.(p\text{ is stationary in }t')\}
$$
If $R$ is a set of transactions, each $t\in R$ over some $Q\subseteq P$ and $S$, then the \temph{$P$-closure of $R$}, $R{\uparrow}P$, is the set of transitions over $P$ and $S$ defined by:
$$
R{\uparrow}P := \bigcup_{t\in R} t{\uparrow}P
$$
Given an equivalence ${\sim}$ on $R$, its \temph{$P$-closure} ${\sim}{\uparrow}P$ is the relation on $R{\uparrow}P$ with $\hat t \mathrel{({\sim}{\uparrow}P)} \hat t'$ iff $\hat t \in t{\uparrow}P$ and $\hat t' \in t'{\uparrow}P$ for some $t \sim t'$.
\end{definition}

Namely, the closure over $P\supseteq Q$ of a transaction $t$ over $Q$ includes all transitions $t'$ over $P$ in which members of $Q$ do the same in $t$ and in $t'$, and the rest remain in their current (arbitrary) state. The closure likewise carries any equivalence on transactions to one on the induced transitions: since distinct transactions over the same participants have disjoint closures, ${\sim}{\uparrow}P$ is an equivalence whenever ${\sim}$ is.

\begin{definition}[Transactions-Based Multiagent Transition System]\label{def:tbmts}
Given agents $P \subset \Pi$, local states $S$ with initial local state $s_0\in S$, a set of transactions $R$, each $t\in R$ over some $Q\subseteq P$ and $S$, and an equivalence ${\sim}$ on $R$, the \temph{transactions-based multiagent transition system} over $P$, $S$, $R$, and ${\sim}$ is the multiagent transition system $TS= (S^P,\{s_0\}^P,R{\uparrow}P,{\sim}{\uparrow}P)$.
\end{definition}

%------------------------------------------------------------------------------
\subsection{From cGLP to maGLP}
\label{sec:glp-to-maglp}\label{sec:maglp-def}
%------------------------------------------------------------------------------

In extending GLP to multiple agents, each agent maintains its own asynchronous resolvent as its local state. The key insight is that GLP's variable pairs provide natural binary communication channels: when agent $p$ assigns a writer $X$ for which the paired reader $X?$ is held by agent $q$, the assignment $X := T$ must be communicated to $q$.

A key difference between cGLP and maGLP is in the initial state. In a multiagent transition system all agents must have the same initial local state $s_0$ (Definition~\ref{def:mts}),  precluding the initial sharing of logic variables, as this would imply different initial states for different agents.
We resolve this in two steps. First, we employ only anonymous logic variables ``\verb|_|'' in the initial local states of agents: Anonymous variables are, on the one hand, syntactically identical, hence allow all initial states to be syntactically identical, and on the other hand represent unique variables, hence semantically all initial goals have unique, local, non-shared variables. The initial state of all agents is the atomic goal \verb|agent(ch(_?,_),ch(_?,_))|, providing two bidirectional channels: the first to the person operating the machine, and the second to the network. The first channel is a bidirectional stream of GLP messages between the machine and the person, mediated by runtime UI support below the GLP layer and by physical UI hardware; messages flowing toward the person are rendered as UI elements, and the person's responses flow back as GLP messages. With appropriate UI runtime support, a GLP message to the user may contain a writer, in which case the runtime presents it as a question and an eventual response from the person becomes the corresponding assignment. The second channel carries communication with other agents.

Second, the Cold-call transaction enables agents to bootstrap communication by establishing shared variables through the network infrastructure, realising the cold-call protocol for connecting previously-disconnected agents.

\begin{definition}[Multiagent GLP]\label{definition:maGLP}
Given agents $P\subset \Pi$ and GLP program $M$, the \temph{maGLP transition system} over $P$ and $M$ is the transactions-based multiagent transition system (Definition~\ref{def:tbmts}) over $P$, local states being asynchronous resolvents $(G_p, \sigma_p)$ over $M$, initial local state $s_0 = (\{\verb|agent(ch(_?,_),ch(_?,_))|\}, \emptyset)$, the following transactions $c\rightarrow c'$, and the equivalence ${\sim}$ given below:
\begin{enumerate}
    \item \textbf{Reduce $p$:} A unary transaction with participant $p$ where $c_p \rightarrow c'_p$ is a cGLP Reduce transition (Definition~\ref{def:cglp-ts}).

    \item \textbf{Communicate $p$ to $q$:} A transaction with participants $p,q\in P$ where $\{X? := T\} \in \sigma_p$, $X?$ occurs in $G_q$, $\sigma'_p = \sigma_p \setminus \{X? := T\}$, and $c'_q = (G_q\{X? := T\}, \sigma_q)$.

    \item \textbf{Cold-call $p$ to $q$:} A binary transaction with participants $p\ne q \in P$ where the network output stream in $c_p$ has a new message \verb|msg|$(q,X)$, $c'_p$ is the result of advancing the network output stream in $c_p$, and $c'_q$ is the result of adding \verb|msg|$(q,X)$ to the network input stream in $c_q$.

    \item ${\sim}$, the \temph{maGLP transaction equivalence} on the transactions above, relates $t_1 \sim t_2$ iff both are Reduce transactions at the same agent $p$ reducing the same goal (by identity, Definition~\ref{def:goal-identity}) with the same clause, both are Communicate transactions from the same $p$ to the same $q$ applying the counterpart of the same writer assignment to the same goal (by identity), or both are Cold-call transactions from the same $p$ to the same $q$ delivering the same message.
\end{enumerate}
\end{definition}

Note that Communicate may be unary or binary, depending on whether $p=q$.  Communicate transfers assignments from writers to readers between agents sharing a paired reader and writer.  Cold-call transfers a term with its variables to $q$ through the network streams established in each agent's initial configuration, enabling the creation of paired variables among previously-disconnected agents.
Cold-call is the exception, as once agents share a paired variable they can use it to communicate indefinitely; moreover, an agent with two friends (with which it shares channels) may introduce them to each other, also eschewing the need for a Cold-call.

%------------------------------------------------------------------------------
\subsection{Safety Properties of maGLP}
\label{sec:maglp-safety}
%------------------------------------------------------------------------------

The safety properties established for cGLP in Section~\ref{sec:glp} extend to maGLP. SO preservation (cf.\ Proposition~\ref{prop:so-preservation}) generalises directly:

\begin{restatable}[maGLP SO Preservation]{proposition}{propMaGLPSOPreservation}
\label{prop:maglp-so-preservation}
If the initial goals of all agents satisfy SO, then every goal in every agent's resolvent throughout a proper run satisfies SO.
\end{restatable}

\begin{definition}[maGLP Proper Run, Outcome]
\label{def:maglp-proper-run}
An maGLP run is \temph{proper} if the run at each agent is proper (Definition~\ref{def:cglp-proper-run}). The \temph{outcome} of a proper maGLP run is the tuple of per-agent outcomes.
\end{definition}

To relate maGLP to deduction, consider the \temph{lifted system} $L$: the cGLP transition system whose resolvent is the union of all agents' local resolvents, whose initial goal includes a \texttt{network} goal with channels paired to each agent's network channels, and whose program $M'$ is $M$ augmented with the GLP definition of \texttt{network}.

\begin{restatable}{proposition}{propMaGLPDeduction}
\label{prop:maglp-deduction}
Every proper maGLP run over $P$ and $M$ is simulated by a proper cGLP run of $L$ whose outcome is a logical consequence of $M'/?$.
\end{restatable}

Every maGLP run is safe, so correctness (Definition~\ref{def:ts}) reduces to liveness.

\mypara{Implementation}
The implementation-ready deterministic variants of GLP and maGLP (the latter deterministic at the agent, not system, level), along with formal correctness proofs that they implement their respective specifications, are presented in a companion paper~\cite{shapiro2026implementing}.

%==============================================================================
\section{The Grassroots Social Graph}
\label{sec:ma-social-graph}
%==============================================================================

The grassroots social graph is the foundation upon which all other grassroots platforms are built. Nodes represent cryptographically self-identified agents; edges represent authenticated bidirectional channels; connected components arise spontaneously through befriending.

Each agent processes messages from user and network input streams and maintains an outputs list with one typed entry per destination --- \verb|user_output|, \verb|net_output|, and a \verb|friend_output| per friend --- on which the library routers \verb|send_user|, \verb|send_net|, and \verb|send_friend| dispatch. The program supports three protocols: (1) \emph{cold-call befriending}, where agents with no prior shared variables establish friendship by exchanging a response variable through the network; (2) \emph{friend-mediated introduction}, where a mutual friend creates a channel pair and sends each half to the respective parties, establishing a direct connection; and (3) \emph{text messaging} between established friends via named output streams.

The core of the program is the procedure \verb|agent/4|; the key clauses follow. Module declarations (\verb|exported|/\verb|imported|, \verb|#|-qualified calls) are elided from quoted clauses.

\mypara{Channels}
A \emph{channel} is an authenticated bidirectional communication structure \verb|ch(In, Out)|, where \verb|In| is a stream of incoming messages and \verb|Out| is a stream of outgoing messages. Two agents share a channel by holding opposite views of the same underlying stream pair: what one writes on its \verb|Out|, the other reads from its \verb|In|. The procedure \verb|new_channel/2| creates such a pair:
\begin{verbatim}
new_channel(ch(Xs?, Ys), ch(Ys?, Xs)).
\end{verbatim}
A single call \verb|new_channel(PQCh, QPCh)| binds \verb|PQCh| to \verb|ch(Xs?, Ys)| and \verb|QPCh| to \verb|ch(Ys?, Xs)|: the reader \verb|Xs?| of one channel is paired with the writer \verb|Xs| of the other (and likewise for \verb|Ys|/\verb|Ys?|), so each end's outgoing stream is the other end's incoming stream. Sending and receiving on a channel are:
\begin{verbatim}
send(X, ch(In, [X?|Out?]), ch(In?, Out)).
receive(X?, ch([X|In], Out?), ch(In?, Out)).
\end{verbatim}
Channels are used in friend-mediated introduction (below) to establish a direct connection between two parties without a cold-call. They are also used by the cold-call protocol: when the recipient accepts, it returns a freshly-created setup-channel half as its response; the friend channel itself is conveyed over the setup channel by the shared commit procedure \verb|befriend_commit|, completing the bidirectional link.

\mypara{Cold-call befriending}
The user's \verb|connect(Target)| request triggers a cold-call: the agent sends \verb|intro(Id, Resp)| on the network and injects the paired reader \verb|Resp?| into its own user-input stream, so the eventual response surfaces locally as a normal user-side message:
\begin{verbatim}
agent(Id, [msg('_user', Id1, connect(Target))|UserIn], NetIn, Outs) :-
    Id? =?= Id1?, ground(Target?) |
    send_net(msg(Target?, intro(Id?, Resp)), Outs?, Outs1),
    inject_msg(Resp?, Target?, Id?, UserIn?, UserIn1),
    agent(Id?, UserIn1?, NetIn?, Outs1?).
\end{verbatim}
On the recipient side, an incoming \verb|intro(From, Resp?)| captures the response writer \verb|Resp|, forwarded to the user in a befriend prompt; the user's decision will assign it, transparently transported back to the originator:
\begin{verbatim}
agent(Id, UserIn, [msg(Id1, intro(From, Resp?))|NetIn], Outs) :-
    Id? =?= Id1? |
    send_user(msg(agent, '_user', befriend(From?, Resp)), Outs?, Outs1),
    agent(Id?, UserIn?, NetIn?, Outs1?).
\end{verbatim}
When the user makes a decision (accept or reject), the agent processes the response:
\begin{verbatim}
agent(Id, [msg('_user', Id1,
    decision(Dec, From, response(Resp?)))|UserIn], NetIn, Outs) :-
    Id? =?= Id1? |
    bind_response(Id?, Dec?, From?, Resp, Outs?, Outs1, NetIn?, NetIn1),
    agent(Id?, UserIn?, NetIn1?, Outs1?).
\end{verbatim}
The helper \verb|bind_response| handles both cases: on \verb|yes|, it creates a fresh setup-channel pair via \verb|new_channel|, assigns \verb|accept(RetCh?)| to \verb|Resp| (transported back to the initiator by the Communicate transaction) and commits the friendship over the local half via the shared \verb|befriend_commit|, which conveys a fresh friend channel over the setup channel, registers its \verb|friend_output| entry, and merges its incoming end into \verb|NetIn|; on \verb|no|, it informs the user the request was rejected.  Both ends run the same idempotent commit.  Its definition appears in \appref{appendix:social-graph-walkthrough}{\arxivref}.

When Alice executes this protocol to befriend Bob, her agent sends \verb|msg(bob, intro(alice, Resp))| on the \verb|'_net'| output stream. The Cold-call transaction (Definition~\ref{definition:maGLP}) transfers this message to Bob's network input stream, adding the writer \verb|Resp| to Bob's resolvent. When Bob accepts, assigning \verb|Resp| to \verb|accept(RetCh?)| for a fresh setup channel, the Communicate transaction transfers this assignment back to Alice; \verb|befriend_commit| then installs the friend channel at both ends.

\mypara{Friend-mediated introduction}
Friend-mediated introduction creates a fresh setup-channel pair via \verb|new_channel/2| and sends one half to each of the introduced parties; on mutual acceptance a direct connection arises without a cold-call:
\begin{verbatim}
agent(Id, [msg('_user', Id1, introduce(P, Q))|UserIn], NetIn, Outs) :-
    Id? =?= Id1?, ground(P?), ground(Q?), ~(P? =?= Q?),
    new_channel(PQCh, QPCh) |
    send_friend(P?, msg(Id?, P?, intro(Q?, QPCh?)), Outs?, Outs1),
    send_friend(Q?, msg(Id?, Q?, intro(P?, PQCh?)), Outs1?, Outs2),
    agent(Id?, UserIn?, NetIn?, Outs2?).
\end{verbatim}
When Bob introduces Alice to Charlie, the setup-channel pair he creates contains readers that are transferred to Alice and Charlie via Communicate; each party accepts by acknowledging on the setup channel, and on mutual acknowledgement both commit the friendship via \verb|befriend_commit| --- a direct connection without any Cold-call.

\mypara{Text messaging}
Once two agents are friends, messages flow directly along their shared channel via the Communicate transaction, without cold-calls.  The user's \verb|send(Target, Text)| request is dispatched by \verb|send_friend| on the \verb|friend_output| entry registered under \verb|Target| in \verb|Outs|:
\begin{verbatim}
agent(Id, [msg('_user', Id1, send(Target, Text))|UserIn], NetIn, Outs) :-
    Id? =?= Id1?, ground(Target?) |
    send_friend(Target?, msg(Id?, Target?, text(Text?)), Outs?, Outs1),
    agent(Id?, UserIn?, NetIn?, Outs1?).
\end{verbatim}
On the other side, an incoming \verb|text(Text)| from \verb|From| is forwarded to the user as \verb|received(From, Text)|:
\begin{verbatim}
agent(Id, UserIn, [msg(From, Id1, text(Text))|NetIn], Outs) :-
    Id? =?= Id1? |
    send_user(msg(agent, '_user', received(From?, Text?)), Outs?, Outs1),
    agent(Id?, UserIn?, NetIn?, Outs1?).
\end{verbatim}

\mypara{Boot and deployment}
Each agent runs on a separate isolate (later --- separate smartphone). A boot clause reduces the universal initial state \verb|agent(ch(_?,_),ch(_?,_))| (Definition~\ref{definition:maGLP}) to the four-argument form used in the clauses above. The complete tested program --- the type vocabulary, helper procedures, the UI mediator, the actors, the plays, and the boot variants --- is the directory \verb|programs/social/graph| of the public GLP repository (\url{https://github.com/EShapiro2/GLP}); a comprehensive Alice/Bob/Charlie scenario appears in \appref{appendix:social-graph-walkthrough}{\arxivref}.

The social graph specified above, run as a maGLP program, is itself a grassroots platform: it inherits the grassroots property of maGLP (Corollary~\ref{corollary:social-graph-grassroots}).

%==============================================================================
\section{Multiagent GLP is Grassroots}\label{sec:grassroots}
%==============================================================================

This section establishes that maGLP is grassroots. We define the notion of grassroots following~\cite{lewis2026volitional}; formal definitions appear in \appref{appendix:grassroots-defs}{\arxivref}.

Informally, a protocol is \emph{grassroots} if two disjoint groups of agents can each operate independently---their interleaved correct runs are correct runs of the combined system---yet the combined system offers genuinely new behaviours that neither group could produce on its own. The first requirement is \emph{obliviousness}; the second is \emph{interactivity}.

The Cold-call transaction (Definition~\ref{definition:maGLP}) is the fundamental interactive transaction of maGLP. It allows an agent $q$ in one group to connect to an agent $p$ in another, establishing shared variables that span both groups---a behaviour that no interleaving of independent runs of the two groups could produce.

\begin{restatable}{theorem}{thmMaGLPGrassroots}
\label{theorem:maGLP-grassroots}
The maGLP protocol is grassroots.
\end{restatable}

The grassroots property of maGLP extends to applications built on top of it, provided they use the cold-call mechanism.

\begin{definition}[GLP Application]
A \temph{GLP application} is a GLP program $M$ together with the maGLP infrastructure. An application \temph{uses cold-calls} if agents can execute the Cold-call transaction to establish communication with previously-disconnected agents.
\end{definition}

\begin{restatable}{proposition}{propAppGrassroots}
\label{prop:app-grassroots}
Any GLP application that uses cold-calls is grassroots.
\end{restatable}

\begin{restatable}{corollary}{corSocialGraphGrassroots}
\label{corollary:social-graph-grassroots}
The GLP implementation of the grassroots social graph (Section~\ref{sec:ma-social-graph}) is grassroots.
\end{restatable}

\section{Conclusion}\label{section:conclusion}

While concurrent logic programming and its powerful programming techniques have been known for four decades~\cite{shapiro1989family}, adoption has been hampered by their inaccessibility to the average programmer. Concurrent logic programming requires a substantial shift from sequential procedural or functional thinking to dataflow programming via partial bindings, with concurrent operational semantics in which communication occurs through binding paired logical variables which may contain further logic variables.  Our experience to date is that AI is highly effective and productive at programming from mathematical specifications when the target is a typed high-level language such as GLP, where types constrain the space of legal programs and provide a checkable specification at the human-AI interface~\cite{shapiro2026types}. Moreover, our preliminary experience suggests that the abstract nature of concurrent logic programming renders it a more powerful and productive language than mainstream languages for collaborative human-AI abstract specifications-based program development.

Constitutional governance of grassroots communities and federations builds on Constitutional Consensus~\cite{keidar2025constitutional,lewis2025morpheus} and digital social contracts~\cite{cardelli2020digital,shapiro2022foundations}, both realisable in GLP.  The GLP runtime and example programs are open-source at \url{https://github.com/EShapiro2/GLP}.

\bibliographystyle{eptcs}
\bibliography{bib}

\begin{thebibliography}{10}
\providecommand{\bibitemdeclare}[2]{}
\providecommand{\surnamestart}{}
\providecommand{\surnameend}{}
\providecommand{\urlprefix}{Available at }
\providecommand{\url}[1]{\texttt{#1}}
\providecommand{\href}[2]{\texttt{#2}}
\providecommand{\urlalt}[2]{\href{#1}{#2}}
\providecommand{\doi}[1]{doi:\urlalt{https://doi.org/#1}{#1}}
\providecommand{\eprint}[1]{arXiv:\urlalt{https://arxiv.org/abs/#1}{#1}}
\providecommand{\bibinfo}[2]{#2}

\bibitemdeclare{article}{armstrong2010erlang}
\bibitem{armstrong2010erlang}
\bibinfo{author}{Joe \surnamestart Armstrong\surnameend}
  (\bibinfo{year}{2010}): \emph{\bibinfo{title}{Erlang}}.
\newblock {\slshape \bibinfo{journal}{Communications of the ACM}}
  \bibinfo{volume}{53}(\bibinfo{number}{9}), pp. \bibinfo{pages}{68--75},
  \doi{10.1145/1810891.1810910}.

\bibitemdeclare{inproceedings}{baker1977future}
\bibitem{baker1977future}
\bibinfo{author}{Henry~G. \surnamestart Baker\surnameend} \&
  \bibinfo{author}{Carl \surnamestart Hewitt\surnameend}
  (\bibinfo{year}{1977}): \emph{\bibinfo{title}{The Incremental Garbage
  Collection of Processes}}.
\newblock In: {\slshape \bibinfo{booktitle}{Proceedings of the 1977 Symposium
  on Artificial Intelligence and Programming Languages}},
  \bibinfo{publisher}{ACM}, pp. \bibinfo{pages}{55--59},
  \doi{10.1145/800228.806932}.

\bibitemdeclare{book}{hansen1973operating}
\bibitem{hansen1973operating}
\bibinfo{author}{Per \surnamestart Brinch~Hansen\surnameend}
  (\bibinfo{year}{1973}): \emph{\bibinfo{title}{Operating System Principles}}.
\newblock \bibinfo{publisher}{Prentice-Hall}.

\bibitemdeclare{inproceedings}{cardelli2020digital}
\bibitem{cardelli2020digital}
\bibinfo{author}{Luca \surnamestart Cardelli\surnameend}, \bibinfo{author}{Liav
  \surnamestart Orgad\surnameend}, \bibinfo{author}{Gal \surnamestart
  Shahaf\surnameend}, \bibinfo{author}{Ehud \surnamestart Shapiro\surnameend}
  \& \bibinfo{author}{Nimrod \surnamestart Talmon\surnameend}
  (\bibinfo{year}{2020}): \emph{\bibinfo{title}{Digital social contracts: A
  foundation for an egalitarian and just digital society}}.
\newblock In: {\slshape \bibinfo{booktitle}{CEUR Proceedings of the First
  International Forum on Digital and Democracy}}, \bibinfo{volume}{2781},
  \bibinfo{organization}{CEUR-WS}, pp. \bibinfo{pages}{51--60}.

\bibitemdeclare{article}{clark1986parlog}
\bibitem{clark1986parlog}
\bibinfo{author}{Keith \surnamestart Clark\surnameend} \&
  \bibinfo{author}{Steve \surnamestart Gregory\surnameend}
  (\bibinfo{year}{1986}): \emph{\bibinfo{title}{PARLOG: parallel programming in
  logic}}.
\newblock {\slshape \bibinfo{journal}{ACM Transactions on Programming Languages
  and Systems (TOPLAS)}} \bibinfo{volume}{8}(\bibinfo{number}{1}), pp.
  \bibinfo{pages}{1--49}, \doi{10.1145/5001.5390}.

\bibitemdeclare{inproceedings}{codish1986compiling}
\bibitem{codish1986compiling}
\bibinfo{author}{Michael \surnamestart Codish\surnameend} \&
  \bibinfo{author}{Ehud \surnamestart Shapiro\surnameend}
  (\bibinfo{year}{1986}): \emph{\bibinfo{title}{Compiling OR-parallelism into
  AND-parallelism}}.
\newblock In: {\slshape \bibinfo{booktitle}{International Conference on Logic
  Programming}}, \bibinfo{organization}{Springer}, pp.
  \bibinfo{pages}{283--297}, \doi{10.1007/3-540-16492-8\_82}.

\bibitemdeclare{misc}{eitan2026securing}
\bibitem{eitan2026securing}
\bibinfo{author}{Ohad \surnamestart Eitan\surnameend}, \bibinfo{author}{Idit
  \surnamestart Keidar\surnameend} \& \bibinfo{author}{Ehud \surnamestart
  Shapiro\surnameend} (\bibinfo{year}{2026}): \emph{\bibinfo{title}{Securing
  People and their Machines Against Major Faults}}.
\newblock \eprint{2607.02304}.

\bibitemdeclare{article}{friedman1976impact}
\bibitem{friedman1976impact}
\bibinfo{author}{Daniel~P. \surnamestart Friedman\surnameend} \&
  \bibinfo{author}{David~S. \surnamestart Wise\surnameend}
  (\bibinfo{year}{1976}): \emph{\bibinfo{title}{The Impact of Applicative
  Programming on Multiprocessing}}.
\newblock {\slshape \bibinfo{journal}{Indiana University Computer Science
  Department Technical Report}} (\bibinfo{number}{TR-26}).

\bibitemdeclare{article}{girard1987linear}
\bibitem{girard1987linear}
\bibinfo{author}{Jean-Yves \surnamestart Girard\surnameend}
  (\bibinfo{year}{1987}): \emph{\bibinfo{title}{Linear Logic}}.
\newblock {\slshape \bibinfo{journal}{Theoretical Computer Science}}
  \bibinfo{volume}{50}(\bibinfo{number}{1}), pp. \bibinfo{pages}{1--101},
  \doi{10.1016/0304-3975(87)90045-4}.

\bibitemdeclare{article}{halpern2024federated}
\bibitem{halpern2024federated}
\bibinfo{author}{Daniel \surnamestart Halpern\surnameend},
  \bibinfo{author}{Ariel~D \surnamestart Procaccia\surnameend},
  \bibinfo{author}{Ehud \surnamestart Shapiro\surnameend} \&
  \bibinfo{author}{Nimrod \surnamestart Talmon\surnameend}
  (\bibinfo{year}{2024}): \emph{\bibinfo{title}{Federated Assemblies}}.
\newblock {\slshape \bibinfo{journal}{Proc AAAI 2025; arXiv preprint
  arXiv:2405.19129}}.

\bibitemdeclare{article}{hoare1974monitors}
\bibitem{hoare1974monitors}
\bibinfo{author}{C.~A.~R. \surnamestart Hoare\surnameend}
  (\bibinfo{year}{1974}): \emph{\bibinfo{title}{Monitors: An operating system
  structuring concept}}.
\newblock {\slshape \bibinfo{journal}{Communications of the ACM}}
  \bibinfo{volume}{17}(\bibinfo{number}{10}), pp. \bibinfo{pages}{549--557},
  \doi{10.1145/355620.361161}.

\bibitemdeclare{article}{houri1989sequential}
\bibitem{houri1989sequential}
\bibinfo{author}{Avshalom \surnamestart Houri\surnameend} \&
  \bibinfo{author}{Ehud \surnamestart Shapiro\surnameend}
  (\bibinfo{year}{1989}): \emph{\bibinfo{title}{A sequential abstract machine
  for Flat Concurrent Prolog}}.
\newblock {\slshape \bibinfo{journal}{The Journal of Logic Programming}}
  \bibinfo{volume}{7}(\bibinfo{number}{2}), pp. \bibinfo{pages}{85--123},
  \doi{10.1016/0743-1066(89)90011-3}.

\bibitemdeclare{article}{keidar2025constitutional}
\bibitem{keidar2025constitutional}
\bibinfo{author}{Idit \surnamestart Keidar\surnameend}, \bibinfo{author}{Andrew
  \surnamestart Lewis-Pye\surnameend} \& \bibinfo{author}{Ehud \surnamestart
  Shapiro\surnameend} (\bibinfo{year}{2025}):
  \emph{\bibinfo{title}{Constitutional Consensus}}.
\newblock {\slshape \bibinfo{journal}{arXiv preprint arXiv:2505.19216}}.

\bibitemdeclare{inproceedings}{kermarrec2020gossiping}
\bibitem{kermarrec2020gossiping}
\bibinfo{author}{Anne-Marie \surnamestart Kermarrec\surnameend},
  \bibinfo{author}{Erick \surnamestart Lavoie\surnameend} \&
  \bibinfo{author}{Christian \surnamestart Tschudin\surnameend}
  (\bibinfo{year}{2020}): \emph{\bibinfo{title}{Gossiping with append-only logs
  in secure-Scuttlebutt}}.
\newblock In: {\slshape \bibinfo{booktitle}{Proceedings of the 1st
  international workshop on distributed infrastructure for common good}}, pp.
  \bibinfo{pages}{19--24}, \doi{10.1145/3428662.3428794}.

\bibitemdeclare{inproceedings}{kowalski1974predicate}
\bibitem{kowalski1974predicate}
\bibinfo{author}{Robert \surnamestart Kowalski\surnameend}
  (\bibinfo{year}{1974}): \emph{\bibinfo{title}{Predicate logic as programming
  language}}.
\newblock In: {\slshape \bibinfo{booktitle}{IFIP congress}},
  \bibinfo{volume}{74}, pp. \bibinfo{pages}{569--574}.

\bibitemdeclare{inproceedings}{levi1985readonly}
\bibitem{levi1985readonly}
\bibinfo{author}{Giorgio \surnamestart Levi\surnameend} \&
  \bibinfo{author}{Catuscia \surnamestart Palamidessi\surnameend}
  (\bibinfo{year}{1985}): \emph{\bibinfo{title}{The Semantics of the Read-Only
  Variable}}.
\newblock In: {\slshape \bibinfo{booktitle}{Proc. Symposium on Logic
  Programming}}, \bibinfo{publisher}{IEEE}, pp. \bibinfo{pages}{128--137}.

\bibitemdeclare{article}{lewis2023grassroots}
\bibitem{lewis2023grassroots}
\bibinfo{author}{Andrew \surnamestart Lewis-Pye\surnameend},
  \bibinfo{author}{Oded \surnamestart Naor\surnameend} \& \bibinfo{author}{Ehud
  \surnamestart Shapiro\surnameend} (\bibinfo{year}{2023}):
  \emph{\bibinfo{title}{Grassroots Flash: A Payment System for Grassroots
  Cryptocurrencies}}.
\newblock {\slshape \bibinfo{journal}{arXiv preprint arXiv:2309.13191}}.

\bibitemdeclare{article}{lewis2025morpheus}
\bibitem{lewis2025morpheus}
\bibinfo{author}{Andrew \surnamestart Lewis-Pye\surnameend} \&
  \bibinfo{author}{Ehud \surnamestart Shapiro\surnameend}
  (\bibinfo{year}{2025}): \emph{\bibinfo{title}{Morpheus Consensus: Excelling
  on trails and autobahns}}.
\newblock {\slshape \bibinfo{journal}{Proc. 29th Conference on Principles of
  Distributed Systems, OPODIS 2025. arXiv preprint arXiv:2502.08465}},
  \doi{10.4230/LIPIcs.OPODIS.2025.35}.

\bibitemdeclare{article}{lewis2026volitional}
\bibitem{lewis2026volitional}
\bibinfo{author}{Andy \surnamestart Lewis-Pye\surnameend} \&
  \bibinfo{author}{Ehud \surnamestart Shapiro\surnameend}
  (\bibinfo{year}{2026}): \emph{\bibinfo{title}{Volitional Multiagent Atomic
  Transactions: Describing People and their Machines}}.
\newblock {\slshape \bibinfo{journal}{arXiv preprint arXiv:2604.25596}}.

\bibitemdeclare{article}{lichtenstein1988concurrent}
\bibitem{lichtenstein1988concurrent}
\bibinfo{author}{Yossi \surnamestart Lichtenstein\surnameend} \&
  \bibinfo{author}{Ehud \surnamestart Shapiro\surnameend}
  (\bibinfo{year}{1988}): \emph{\bibinfo{title}{Concurrent algorithmic
  debugging}}.
\newblock {\slshape \bibinfo{journal}{ACM SIGPLAN Notices}}
  \bibinfo{volume}{24}(\bibinfo{number}{1}), pp. \bibinfo{pages}{248--260},
  \doi{10.1145/68210.69239}.

\bibitemdeclare{misc}{akka2022}
\bibitem{akka2022}
\bibinfo{author}{\surnamestart {Lightbend Inc.}\surnameend}
  (\bibinfo{year}{2022}): \emph{\bibinfo{title}{Akka: Build Concurrent,
  Distributed, and Resilient Message-Driven Applications}}.
\newblock \bibinfo{howpublished}{\url{https://akka.io}}.
\newblock \bibinfo{note}{Accessed October 2025}.

\bibitemdeclare{book}{lloyd1987foundations}
\bibitem{lloyd1987foundations}
\bibinfo{author}{John~W. \surnamestart Lloyd\surnameend}
  (\bibinfo{year}{1987}): \emph{\bibinfo{title}{Foundations of Logic
  Programming}}, \bibinfo{edition}{2nd} edition.
\newblock \bibinfo{publisher}{Springer-Verlag},
  \doi{10.1007/978-3-642-83189-8}.

\bibitemdeclare{misc}{orleans2022}
\bibitem{orleans2022}
\bibinfo{author}{\surnamestart {Microsoft}\surnameend} (\bibinfo{year}{2022}):
  \emph{\bibinfo{title}{Orleans: Cloud Native Application Framework}}.
\newblock \bibinfo{howpublished}{\url{https://dotnet.github.io/orleans}}.
\newblock \bibinfo{note}{Accessed October 2025}.

\bibitemdeclare{article}{mierowsky1985fcp}
\bibitem{mierowsky1985fcp}
\bibinfo{author}{C.~\surnamestart Mierowsky\surnameend},
  \bibinfo{author}{S.~\surnamestart Taylor\surnameend},
  \bibinfo{author}{E.~\surnamestart Shapiro\surnameend},
  \bibinfo{author}{J.~\surnamestart Levy\surnameend} \&
  \bibinfo{author}{M.~\surnamestart Safra\surnameend} (\bibinfo{year}{1985}):
  \emph{\bibinfo{title}{On the implementation of Flat Concurrent Prolog}}.
\newblock {\slshape \bibinfo{journal}{Proceedings of the 1985 Symposium on
  Logic Programming}}, pp. \bibinfo{pages}{276--286}.

\bibitemdeclare{inproceedings}{moto1983overview}
\bibitem{moto1983overview}
\bibinfo{author}{Tohru \surnamestart Moto-Oka\surnameend}
  (\bibinfo{year}{1983}): \emph{\bibinfo{title}{Overview to the fifth
  generation computer system project}}.
\newblock In: {\slshape \bibinfo{booktitle}{Proceedings of the 10th annual
  international symposium on Computer architecture}}, pp.
  \bibinfo{pages}{417--422}.

\bibitemdeclare{incollection}{safra1988meta}
\bibitem{safra1988meta}
\bibinfo{author}{Shmuel \surnamestart Safra\surnameend} \&
  \bibinfo{author}{Ehud \surnamestart Shapiro\surnameend}
  (\bibinfo{year}{1988}): \emph{\bibinfo{title}{Meta interpreters for real}}.
\newblock In: {\slshape \bibinfo{booktitle}{Concurrent Prolog: Collected
  Papers}}, \bibinfo{publisher}{MIT Press}, pp. \bibinfo{pages}{166--179}.

\bibitemdeclare{incollection}{shafrir1988distributed}
\bibitem{shafrir1988distributed}
\bibinfo{author}{A.~\surnamestart Shafrir\surnameend} \&
  \bibinfo{author}{E.~\surnamestart Shapiro\surnameend} (\bibinfo{year}{1987}):
  \emph{\bibinfo{title}{Distributed Programming in {C}oncurrent {P}rolog}}.
\newblock In \bibinfo{editor}{Ehud \surnamestart Shapiro\surnameend}, editor:
  {\slshape \bibinfo{booktitle}{Concurrent Prolog: Collected Papers, Volume
  1}}, \bibinfo{publisher}{MIT Press}, \bibinfo{address}{Cambridge, MA}, pp.
  \bibinfo{pages}{318--338}.

\bibitemdeclare{article}{shapiro1983fifth}
\bibitem{shapiro1983fifth}
\bibinfo{author}{Ehud \surnamestart Shapiro\surnameend} (\bibinfo{year}{1983}):
  \emph{\bibinfo{title}{The fifth generation project—a trip report}}.
\newblock {\slshape \bibinfo{journal}{Communications of the ACM}}
  \bibinfo{volume}{26}(\bibinfo{number}{9}), pp. \bibinfo{pages}{637--641},
  \doi{10.1145/358172.358179}.

\bibitemdeclare{article}{shapiro1983subset}
\bibitem{shapiro1983subset}
\bibinfo{author}{Ehud \surnamestart Shapiro\surnameend} (\bibinfo{year}{1983}):
  \emph{\bibinfo{title}{A subset of Concurrent Prolog and its interpreter}}.
\newblock {\slshape \bibinfo{journal}{ICOT Technical Report, TR-003}}.

\bibitemdeclare{inproceedings}{shapiro1984systems}
\bibitem{shapiro1984systems}
\bibinfo{author}{Ehud \surnamestart Shapiro\surnameend} (\bibinfo{year}{1984}):
  \emph{\bibinfo{title}{Systems programming in concurrent prolog}}.
\newblock In: {\slshape \bibinfo{booktitle}{Proceedings of the 11th ACM
  SIGACT-SIGPLAN symposium on Principles of Programming Languages}}, pp.
  \bibinfo{pages}{93--105}, \doi{10.1145/800017.800520}.

\bibitemdeclare{book}{shapiro1987concurrent}
\bibitem{shapiro1987concurrent}
\bibinfo{author}{Ehud \surnamestart Shapiro\surnameend} (\bibinfo{year}{1987}):
  \emph{\bibinfo{title}{Concurrent Prolog: collected papers (Vols. I and II)}}.
\newblock \bibinfo{publisher}{MIT press}.

\bibitemdeclare{article}{shapiro1989family}
\bibitem{shapiro1989family}
\bibinfo{author}{Ehud \surnamestart Shapiro\surnameend} (\bibinfo{year}{1989}):
  \emph{\bibinfo{title}{The family of concurrent logic programming languages}}.
\newblock {\slshape \bibinfo{journal}{ACM Computing Surveys (CSUR)}}
  \bibinfo{volume}{21}(\bibinfo{number}{3}), pp. \bibinfo{pages}{413--510},
  \doi{10.1145/72551.72555}.

\bibitemdeclare{article}{shapiro1989or}
\bibitem{shapiro1989or}
\bibinfo{author}{Ehud \surnamestart Shapiro\surnameend} (\bibinfo{year}{1989}):
  \emph{\bibinfo{title}{Or-parallel prolog in flat concurrent prolog}}.
\newblock {\slshape \bibinfo{journal}{The Journal of Logic Programming}}
  \bibinfo{volume}{6}(\bibinfo{number}{3}), pp. \bibinfo{pages}{243--267},
  \doi{10.1016/0743-1066(89)90016-2}.

\bibitemdeclare{article}{shapiro2021multiagent}
\bibitem{shapiro2021multiagent}
\bibinfo{author}{Ehud \surnamestart Shapiro\surnameend} (\bibinfo{year}{2021}):
  \emph{\bibinfo{title}{Multiagent Transition Systems: Protocol-Stack
  Mathematics for Distributed Computing}}.
\newblock {\slshape \bibinfo{journal}{arXiv preprint arXiv:2112.13650}}.

\bibitemdeclare{inproceedings}{shapiro2023grassrootsBA}
\bibitem{shapiro2023grassrootsBA}
\bibinfo{author}{Ehud \surnamestart Shapiro\surnameend} (\bibinfo{year}{2023}):
  \emph{\bibinfo{title}{Grassroots Distributed Systems: Concept, Examples,
  Implementation and Applications (Brief Announcement)}}.
\newblock In: {\slshape \bibinfo{booktitle}{37th International Symposium on
  Distributed Computing (DISC 2023). (Extended version: arXiv:2301.04391)}},
  \bibinfo{publisher}{LIPICS}, \bibinfo{address}{Italy}, pp.
  \bibinfo{pages}{47:1, 47:7}.

\bibitemdeclare{inproceedings}{shapiro2023gsn}
\bibitem{shapiro2023gsn}
\bibinfo{author}{Ehud \surnamestart Shapiro\surnameend} (\bibinfo{year}{2023}):
  \emph{\bibinfo{title}{Grassroots Social Networking: Serverless,
  Permissionless Protocols for Twitter/LinkedIn/WhatsApp}}.
\newblock In: {\slshape \bibinfo{booktitle}{OASIS ’23}},
  \bibinfo{publisher}{Association for Computing Machinery},
  \doi{10.1145/3599696.3612898}.

\bibitemdeclare{article}{shapiro2024gc}
\bibitem{shapiro2024gc}
\bibinfo{author}{Ehud \surnamestart Shapiro\surnameend} (\bibinfo{year}{2024}):
  \emph{\bibinfo{title}{Grassroots Currencies: Foundations for Grassroots
  Digital Economies}}.
\newblock {\slshape \bibinfo{journal}{arXiv preprint arXiv:2202.05619}}.

\bibitemdeclare{article}{shapiro2025characterising}
\bibitem{shapiro2025characterising}
\bibinfo{author}{Ehud \surnamestart Shapiro\surnameend} (\bibinfo{year}{2025}):
  \emph{\bibinfo{title}{Characterising Global Platforms: Centralised,
  Decentralised, Federated, and Grassroots}}.
\newblock {\slshape \bibinfo{journal}{arXiv preprint arXiv:2511.03286}}.

\bibitemdeclare{article}{shapiro2025glp}
\bibitem{shapiro2025glp}
\bibinfo{author}{Ehud \surnamestart Shapiro\surnameend} (\bibinfo{year}{2025}):
  \emph{\bibinfo{title}{GLP: A Grassroots, Multiagent, Concurrent, Logic
  Programming Language for AI (full version)}}.
\newblock {\slshape \bibinfo{journal}{arXiv preprint arXiv:2510.15747, Summary
  to appear in Proc. of ICLP'26}}.

\bibitemdeclare{article}{shapiro2026cssn}
\bibitem{shapiro2026cssn}
\bibinfo{author}{Ehud \surnamestart Shapiro\surnameend} (\bibinfo{year}{2026}):
  \emph{\bibinfo{title}{Child-Safe Social Networking}}.
\newblock {\slshape \bibinfo{journal}{Submitted}}.

\bibitemdeclare{misc}{shapiro2026bonds}
\bibitem{shapiro2026bonds}
\bibinfo{author}{Ehud \surnamestart Shapiro\surnameend} (\bibinfo{year}{2026}):
  \emph{\bibinfo{title}{Grassroots Bonds as a Foundation for Market
  Liquidity}}.
\newblock \eprint{2603.13671}.

\bibitemdeclare{inproceedings}{shapiro2025atomic}
\bibitem{shapiro2025atomic}
\bibinfo{author}{Ehud \surnamestart Shapiro\surnameend} (\bibinfo{year}{2026}):
  \emph{\bibinfo{title}{Grassroots Platforms with Atomic Transactions: Social
  Graphs, Cryptocurrencies, and Democratic Federations}}.
\newblock In: {\slshape \bibinfo{booktitle}{Proceedings of the 27th
  International Conference on Distributed Computing and Networking}}, pp.
  \bibinfo{pages}{71--81}, \doi{10.1145/3772290.3772309}.
\newblock \bibinfo{note}{ArXiv preprint arXiv:2502.11299}.

\bibitemdeclare{article}{shapiro2026implementing}
\bibitem{shapiro2026implementing}
\bibinfo{author}{Ehud \surnamestart Shapiro\surnameend} (\bibinfo{year}{2026}):
  \emph{\bibinfo{title}{Implementing Grassroots Logic Programs with Multiagent
  Transition Systems and AI (Full Version)}}.
\newblock {\slshape \bibinfo{journal}{arXiv:2602.06934. Summary to appear in
  Proc. of LOPSTR+PPDP'26}}.

\bibitemdeclare{article}{shapiro2026types}
\bibitem{shapiro2026types}
\bibinfo{author}{Ehud \surnamestart Shapiro\surnameend} (\bibinfo{year}{2026}):
  \emph{\bibinfo{title}{Types for Grassroots Logic Programs (Full version)}}.
\newblock {\slshape \bibinfo{journal}{arXiv preprint arXiv:2601.17957}}.

\bibitemdeclare{article}{shapiro1984fair}
\bibitem{shapiro1984fair}
\bibinfo{author}{Ehud \surnamestart Shapiro\surnameend} \&
  \bibinfo{author}{Colin \surnamestart Mierowsky\surnameend}
  (\bibinfo{year}{1984}): \emph{\bibinfo{title}{Fair, biased, and
  self-balancing merge operators: Their specification and implementation in
  Concurrent Prolog}}.
\newblock {\slshape \bibinfo{journal}{New Generation Computing}}
  \bibinfo{volume}{2}(\bibinfo{number}{3}), pp. \bibinfo{pages}{221--240},
  \doi{10.1007/BF03037058}.

\bibitemdeclare{article}{shapiro1986multiway}
\bibitem{shapiro1986multiway}
\bibinfo{author}{Ehud \surnamestart Shapiro\surnameend} \&
  \bibinfo{author}{Shmuel \surnamestart Safra\surnameend}
  (\bibinfo{year}{1986}): \emph{\bibinfo{title}{Multiway merge with constant
  delay in Concurrent Prolog}}.
\newblock {\slshape \bibinfo{journal}{New Generation Computing}}
  \bibinfo{volume}{4}(\bibinfo{number}{2}), pp. \bibinfo{pages}{211--216},
  \doi{10.1007/BF03037442}.

\bibitemdeclare{article}{shapiro1983object}
\bibitem{shapiro1983object}
\bibinfo{author}{Ehud \surnamestart Shapiro\surnameend} \&
  \bibinfo{author}{Akikazu \surnamestart Takeuchi\surnameend}
  (\bibinfo{year}{1983}): \emph{\bibinfo{title}{Object oriented programming in
  Concurrent Prolog}}.
\newblock {\slshape \bibinfo{journal}{New Generation Computing}}
  \bibinfo{volume}{1}(\bibinfo{number}{1}), pp. \bibinfo{pages}{25--48},
  \doi{10.1007/BF03037020}.

\bibitemdeclare{inproceedings}{shapiro2022foundations}
\bibitem{shapiro2022foundations}
\bibinfo{author}{Ehud \surnamestart Shapiro\surnameend} \&
  \bibinfo{author}{Nimrod \surnamestart Talmon\surnameend}
  (\bibinfo{year}{2022}): \emph{\bibinfo{title}{Foundations for Grassroots
  Democratic Metaverse}}.
\newblock In: {\slshape \bibinfo{booktitle}{Proceedings of the 21st
  International Conference on Autonomous Agents and Multiagent Systems}},
  \bibinfo{series}{AAMAS '22}, \bibinfo{publisher}{International Foundation for
  Autonomous Agents and Multiagent Systems}, \bibinfo{address}{Richland, SC},
  p. \bibinfo{pages}{1814–1818}, \doi{10.65109/cnal9987}.

\bibitemdeclare{article}{shapiro19935th}
\bibitem{shapiro19935th}
\bibinfo{author}{Ehud \surnamestart Shapiro\surnameend} \&
  \bibinfo{author}{David~HD \surnamestart Warren\surnameend}
  (\bibinfo{year}{1993}): \emph{\bibinfo{title}{The 5th Generation Project:
  personal perspectives.}}
\newblock {\slshape \bibinfo{journal}{Communications of the ACM}}
  \bibinfo{volume}{36}(\bibinfo{number}{3}), pp. \bibinfo{pages}{47--49}.

\bibitemdeclare{incollection}{silverman1988logix}
\bibitem{silverman1988logix}
\bibinfo{author}{William \surnamestart Silverman\surnameend},
  \bibinfo{author}{Michael \surnamestart Hirsch\surnameend},
  \bibinfo{author}{Avshalom \surnamestart Houri\surnameend} \&
  \bibinfo{author}{Ehud \surnamestart Shapiro\surnameend}
  (\bibinfo{year}{1988}): \emph{\bibinfo{title}{The Logix system user manual
  Version 1.21}}.
\newblock In: {\slshape \bibinfo{booktitle}{Concurrent Prolog: Collected
  Papers}}, pp. \bibinfo{pages}{46--77}.

\bibitemdeclare{incollection}{takeuchi1988bounded}
\bibitem{takeuchi1988bounded}
\bibinfo{author}{A.~\surnamestart Takeuchi\surnameend} \&
  \bibinfo{author}{K.~\surnamestart Furukawa\surnameend}
  (\bibinfo{year}{1987}): \emph{\bibinfo{title}{Bounded Buffer Communication in
  {C}oncurrent {P}rolog}}.
\newblock In \bibinfo{editor}{Ehud \surnamestart Shapiro\surnameend}, editor:
  {\slshape \bibinfo{booktitle}{Concurrent Prolog: Collected Papers, Volume
  1}}, \bibinfo{publisher}{MIT Press}, \bibinfo{address}{Cambridge, MA}, pp.
  \bibinfo{pages}{464--475}.

\bibitemdeclare{article}{shapiro2025GF}
\bibitem{shapiro2025GF}
\bibinfo{author}{Nimrod \surnamestart Talmon\surnameend} \&
  \bibinfo{author}{Ehud \surnamestart Shapiro\surnameend}
  (\bibinfo{year}{2025}): \emph{\bibinfo{title}{Grassroots Federation: Fair
  Democratic Governance at Scale}}.
\newblock {\slshape \bibinfo{journal}{arXiv preprint arXiv:2505.02208; also
  Proc. of AAMAS'26}}.

\bibitemdeclare{incollection}{tribble1988channels}
\bibitem{tribble1988channels}
\bibinfo{author}{E.~D. \surnamestart Tribble\surnameend},
  \bibinfo{author}{M.~S. \surnamestart Miller\surnameend},
  \bibinfo{author}{K.~\surnamestart Kahn\surnameend}, \bibinfo{author}{D.~G.
  \surnamestart Bobrow\surnameend}, \bibinfo{author}{C.~\surnamestart
  Abbott\surnameend} \& \bibinfo{author}{E.~\surnamestart Shapiro\surnameend}
  (\bibinfo{year}{1987}): \emph{\bibinfo{title}{Channels: A Generalization of
  Streams}}.
\newblock In \bibinfo{editor}{Ehud \surnamestart Shapiro\surnameend}, editor:
  {\slshape \bibinfo{booktitle}{Concurrent Prolog: Collected Papers, Volume
  1}}, \bibinfo{publisher}{MIT Press}, \bibinfo{address}{Cambridge, MA}, pp.
  \bibinfo{pages}{446--463}.

\bibitemdeclare{misc}{Ubique}
\bibitem{Ubique}
\bibinfo{author}{\surnamestart Ubique\surnameend} (\bibinfo{year}{1994}):
  \emph{\bibinfo{title}{Ubique. Wikipedia, The Free Encyclopedia}}.
\newblock \urlprefix\url{https://en.wikipedia.org/wiki/Ubique_(company)}.

\bibitemdeclare{inproceedings}{ueda1986guarded}
\bibitem{ueda1986guarded}
\bibinfo{author}{Kazunori \surnamestart Ueda\surnameend}
  (\bibinfo{year}{1986}): \emph{\bibinfo{title}{Guarded Horn Clauses}}.
\newblock In: {\slshape \bibinfo{booktitle}{Logic Programming '85}}, {\slshape
  \bibinfo{series}{Lecture Notes in Computer Science}} \bibinfo{volume}{221},
  \bibinfo{publisher}{Springer}, pp. \bibinfo{pages}{168--179},
  \doi{10.1007/3-540-16479-0\_17}.

\bibitemdeclare{article}{ueda2001resource}
\bibitem{ueda2001resource}
\bibinfo{author}{Kazunori \surnamestart Ueda\surnameend}
  (\bibinfo{year}{2001}): \emph{\bibinfo{title}{Resource-passing concurrent
  programming}}.
\newblock {\slshape \bibinfo{journal}{Proceedings of TACS 2001}}, pp.
  \bibinfo{pages}{95--126}, \doi{10.1007/3-540-45500-0\_5}.

\bibitemdeclare{article}{ueda1994moded}
\bibitem{ueda1994moded}
\bibinfo{author}{Kazunori \surnamestart Ueda\surnameend} \&
  \bibinfo{author}{Masao \surnamestart Morita\surnameend}
  (\bibinfo{year}{1994}): \emph{\bibinfo{title}{Moded Flat GHC and Its
  Message-Oriented Implementation Technique}}.
\newblock {\slshape \bibinfo{journal}{New Generation Computing}}
  \bibinfo{volume}{12}(\bibinfo{number}{4}), pp. \bibinfo{pages}{337--368},
  \doi{10.1007/BF03038307}.

\bibitemdeclare{inproceedings}{ueda1995io}
\bibitem{ueda1995io}
\bibinfo{author}{Kazunori \surnamestart Ueda\surnameend} \&
  \bibinfo{author}{Masao \surnamestart Morita\surnameend}
  (\bibinfo{year}{1995}): \emph{\bibinfo{title}{I/O mode analysis in concurrent
  logic programming}}.
\newblock In: {\slshape \bibinfo{booktitle}{Proceedings of the International
  Symposium on Theory and Practice of Parallel Programming}},
  \bibinfo{publisher}{Springer}, pp. \bibinfo{pages}{356--368},
  \doi{10.1007/BFb0026579}.

\end{thebibliography}

\ifappendix
\appendix

\input{sections/appendix-lp}
\input{sections/appendix-term-matching}
\input{sections/appendix-grassroots-defs}
\input{sections/appendix-guards}
\input{sections/appendix-proofs}
\input{sections/appendix-social-graph-walkthrough}
\fi

\end{document}